\documentclass[a4paper, amsfonts, amssymb, amsmath, reprint, showkeys, nofootinbib, twoside, superscriptaddress]{revtex4-1}
\usepackage[english]{babel}
\usepackage[utf8]{inputenc}
\usepackage[colorinlistoftodos, color=green!40, prependcaption]{todonotes}
\usepackage{amsthm}
\usepackage{mathtools}

\usepackage{xcolor}
\usepackage{graphicx}
\usepackage[left=23mm,right=13mm,top=35mm,columnsep=15pt]{geometry} 
\usepackage{adjustbox}
\usepackage{placeins}
\usepackage[T1]{fontenc}
\usepackage{lipsum}
\usepackage{csquotes}

\usepackage{amsfonts}
\usepackage{physics}
\usepackage[pdftex, pdftitle={Article}, pdfauthor={Author}]{hyperref} 
\begin{document}

\title{Leveraging Randomized Compiling for the QITE Algorithm}


\author{Jean-Loup Ville}
\thanks{These two authors contributed equally.\\ Correspondence should be adressed to \href{mailto:amorvan@lbl.gov}{amorvan@lbl.gov} and \href{mailto:jlville@berkeley.edu}{jlville@berkeley.edu}}
\affiliation{Quantum Nanoelectronics Laboratory, Dept. of Physics, University of California at Berkeley, Berkeley, CA 94720, USA}

\author{Alexis Morvan}
\thanks{These two authors contributed equally.\\ Correspondence should be adressed to \href{mailto:amorvan@lbl.gov}{amorvan@lbl.gov} and \href{mailto:jlville@berkeley.edu}{jlville@berkeley.edu}}
\affiliation{Quantum Nanoelectronics Laboratory, Dept. of Physics, University of California at Berkeley, Berkeley, CA 94720, USA}
\affiliation{Computational Research Division, Lawrence Berkeley National Lab, Berkeley, CA 94720, USA}

\author{Akel Hashim}
\affiliation{Quantum Nanoelectronics Laboratory, Dept. of Physics, University of California at Berkeley, Berkeley, CA 94720, USA}
\author{Ravi K. Naik}
\affiliation{Quantum Nanoelectronics Laboratory, Dept. of Physics, University of California at Berkeley, Berkeley, CA 94720, USA}
\author{Marie Lu}
\affiliation{Quantum Nanoelectronics Laboratory, Dept. of Physics, University of California at Berkeley, Berkeley, CA 94720, USA}
\author{Bradley Mitchell}
\affiliation{Quantum Nanoelectronics Laboratory, Dept. of Physics, University of California at Berkeley, Berkeley, CA 94720, USA}
\author{John-Mark Kreikebaum}
\affiliation{Quantum Nanoelectronics Laboratory, Dept. of Physics, University of California at Berkeley, Berkeley, CA 94720, USA}
\affiliation{Materials Sciences Division, Lawrence Berkeley National Lab, Berkeley, CA 94720, USA}

\author{Kevin P. O'Brien}
\affiliation{Department of Electrical Engineering and Computer Science, Massachusetts Institute of Technology, Cambridge, MA 02139, USA}

\author{Joel J. Wallman}
\affiliation{Inst. for Quantum Computing and Dept. of Applied Mathematics, University of Waterloo, Waterloo, Ontario N2L 3G1, Canada}
\affiliation{Quantum Benchmark Inc., Kitchener, ON N2H 5G5, Canada}
\author{Ian Hincks}
\affiliation{Quantum Benchmark Inc., Kitchener, ON N2H 5G5, Canada}
\author{Joseph Emerson}
\affiliation{Inst. for Quantum Computing and Dept. of Applied Mathematics, University of Waterloo, Waterloo, Ontario N2L 3G1, Canada}
\affiliation{Quantum Benchmark Inc., Kitchener, ON N2H 5G5, Canada}

\author{Ethan Smith}
\affiliation{Computational Research Division, Lawrence Berkeley National Lab, Berkeley, CA 94720, USA}
\author{Ed Younis}
\affiliation{Computational Research Division, Lawrence Berkeley National Lab, Berkeley, CA 94720, USA}
\author{Costin Iancu}
\affiliation{Computational Research Division, Lawrence Berkeley National Lab, Berkeley, CA 94720, USA}

\author{David I. Santiago}
\affiliation{Quantum Nanoelectronics Laboratory, Dept. of Physics, University of California at Berkeley, Berkeley, CA 94720, USA}
\affiliation{Computational Research Division, Lawrence Berkeley National Lab, Berkeley, CA 94720, USA}

\author{Irfan Siddiqi}
\affiliation{Quantum Nanoelectronics Laboratory, Dept. of Physics, University of California at Berkeley, Berkeley, CA 94720, USA}
\affiliation{Computational Research Division, Lawrence Berkeley National Lab, Berkeley, CA 94720, USA}
\affiliation{Materials Sciences Division, Lawrence Berkeley National Lab, Berkeley, CA 94720, USA}

\date{\today} 

\begin{abstract}
The success of the current generation of Noisy Intermediate-Scale Quantum (NISQ) hardware shows that quantum hardware may be able to tackle complex problems even without error correction. One outstanding issue is that of coherent errors arising from the increased complexity of these devices. These errors can accumulate through a circuit, making their impact on algorithms hard to predict and mitigate. Iterative algorithms like Quantum Imaginary Time Evolution are susceptible to these errors. This article presents the combination of both noise tailoring using Randomized Compiling and error mitigation with a purification. We also show that Cycle Benchmarking gives an estimate of the reliability of the purification. We apply this method to the Quantum Imaginary Time Evolution of a Transverse Field Ising Model and report an energy estimation and a ground state infidelity both below 1\%. Our methodology is general and can be used for other algorithms and platforms. We show how combining noise tailoring and error mitigation will push forward the performance of NISQ devices.
\end{abstract}


\maketitle

\section{Introduction}

To realize impactful application of Noisy Intermediate-Scale Quantum (NISQ) devices, error mitigation strategies have emerged as a principle focus of quantum information science. Unlike quantum error correction, which corrects errors as they occur, error mitigation uses post-processing techniques to reduce the impact of errors on the results of an algorithm. These error mitigation schemes are needed to tackle the noise and errors present in current quantum hardware. Many recently implemented algorithms have required some form of error mitigation with already state-of-the-art hardware \cite{kandala2017hardware, kandala2019error, google2020hartree, dumitrescu2018cloud}. Several types of error mitigation can be distinguished: error extrapolation purposely scales the rate of a specific known error in order to extrapolate the results to zero noise \cite{ying2017, temme2017, dumitrescu2018cloud, kandala2019error} at the expense of additional measurements, and assumptions on the noise. Inverting error protocols characterize errors to then correct them with quasi-probabilities \cite{temme2017, zhang2020}, requiring a precise and extensive characterization of the system using Quantum Process Tomography \cite{Chuang_1997} or Gate Set Tomography \cite{Blume_Kohout_2017}. Post-selection protocols eliminate wrong output solutions by checking, for example, an expected symmetry \cite{mcardle2019, bonet2018}. Such post-selection techniques usually require extra quantum resources, such as ancillary qubits. Each technique requires a careful consideration of the additional measurement overhead required, especially as the applications scale in scope. 

For quantum algorithms on NISQ hardware, one of the biggest challenges comes from coherent errors. Contrary to decoherence, the accumulation of coherent errors strongly depends on the circuit used. These errors arise due to the increasing complexity of quantum devices and originate from multiple mechanisms like crosstalk, frequency collision, drift, etc., making it difficult to track and compensate for their impact. On the contrary, decoherence processes are easier to predict and correct as their behavior does not depend on the circuit. To tackle the problem of coherent errors while keeping the same number of measurements - or total number of shots per experiment - it is possible to tailor them into stochastic noise using Twirling properties, where two-qubit gates are sandwiched between twirling gates, and then statistical averaging. Twirling is a technique now widely known in the quantum information literature. It is for example at the heart of Randomized Benchmarking \cite{Emerson_2005, Knill2008-rz, Magesan_2011, Magesan2012-dy}. Randomized Compiling (RC) \cite{Wallman2016-dn} - that uses Pauli twirling - has been shown to improve the performance of quantum devices \cite{Ware_2021, hashim2020randomized}. 

Common benchmarks of error mitigation techniques on NISQ devices are usually fixed-depth algorithms such as VQE \cite{peruzzo2014variational, mccaskey2019quantum, colless2018computation} or QAOA \cite{farhi2014quantum,harrigan2021quantum,lacroix2020improving}. Here we use the Quantum Imaginary Time Evolution (QITE) \cite{motta2020determining, yeter2019practical, aydeniz2021scattering, McArdle2019-ah} algorithm to benchmark the different error mitigation techniques. It is an iterative algorithm that approximates imaginary time evolution with a unitary operation. In the limit of long imaginary time, the algorithm reaches the ground state of a given Hamiltonian. One advantage of QITE is that it generalizes easily to calculate finite temperature quantities \cite{Sun2020-jw, bassman2021computing}. Additionally it does not require \textit{a priori} knowledge of an ansatz, which can be difficult for variational algorithms \cite{cerezo2020variational, grimsley2019adaptive, choquette2020quantum}. The sensitivity to experimental errors of the computation of each step makes the algorithm less resilient to noise than VQE \cite{omalley2016,McClean2017}, which makes it a good candidate to benchmark error mitigation protocols.

In this article, we use 3 qubits (labelled 4,5 and 6) in a linear topology out of an 8 fixed-frequency transmons chip described in \cite{blok2021quantum, hashim2020randomized}. The single qubit gates are performed using the ZXZXZ decomposition with virtual Z gates \cite{mckay2017efficient}. Our entanglement is based on the cross-resonance interaction and realizes a CZ gate \cite{mitchell2021}. We combine noise tailoring with RC and mitigation with purification and show that using both improves the quality of the result beyond what one would expect from using each one separately. We attribute this performance improvement to the noise tailoring by RC that effectively maps the coherent errors into Pauli errors, which are simpler to handle and can be further approximated as a fully depolarizing error model. In Section \ref{sec:noise} we discuss the implementation of RC with purification and give an estimation of how close the noise is to fully depolarizing using Cycle Benchmarking \cite{erhard2019characterizing} and compare it to our hardware. In Section \ref{sec:qite}, we then use this scheme to perform QITE on the Transverse Ising Field Model with 3 qubits to benchmark the efficacy of this method. We conclude by how to extend and complement our techniques with further mitigation schemes.

\section{Noise tailoring with Randomized compiling} \label{sec:noise}

\begin{figure*}[ht!]
    \centering
    \includegraphics{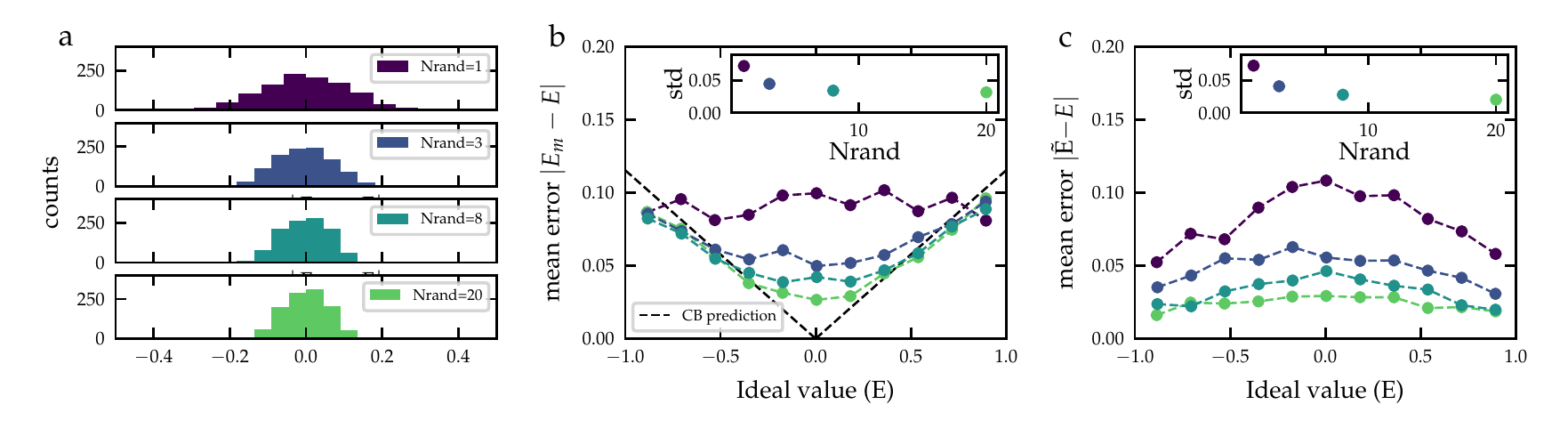}
    \caption{Effect of randomized compiling on measured expectation values for random circuits using 6 CZs. Panel a. shows the distribution of errors on the measured expectation values $|E_m - E|$ for the different numbers of randomizations. Panel b. shows the same data, but sorted by measured expectation values, to show the increasing agreement with a depolarization model $E_m = \bar{\lambda} E$, with $\bar{\lambda}$ the mean of the Pauli decays measured by CB ($\bar{\lambda}_{CB} = 0.980$) --- performed prior to the experiment. The standard deviation of the errors is similar for all points of same number of randomizations and is plotted in the inset. The data in c. shows the reduction of errors for all expectation values when using the purification formula \ref{eq:rescaling}. The standard deviation of the Pauli decays is $\text{std}(\lambda_P) = 0.003$, as shown at the top of Figure~\ref{fig:lambdaP_2}.
    }
    \label{fig:v_shape}
\end{figure*}

Twirling is a powerful technique that tailors the noise a circuit experiences when run on hardware. Its most notable use is the characterization of quantum processors with Randomized Benchmarking (RB) for qubits \cite{Emerson_2005, Knill2008-rz, Magesan_2011, Magesan2012-dy} and qudits \cite{morvan2020qutrit, Jafarzadeh_2020}. Defining a twirl requires a twirling group $\mathcal{G}$ that is usually the Pauli group or the Clifford group. For every cycle of the circuit - determined by the native gate-set - a gate from the twirl group is inserted. Twirling in the mathematical sense is achieved by averaging the circuit outcomes for all combinations of twirling gates. In practice, averaging over a few randomly sampled twirls is enough. When the twirling group is a single qubit group like the Pauli group $\mathcal{P}$, this step is done efficiently by compiling the twirling gate into the circuit's single qubit gates, keeping the circuit's depth unchanged in terms of multi-qubit gates. For twirling to be useful for a specific circuit, one has to track all the twirl operations and invert it at the end of the circuit. Even for a simple twirling group like the Pauli group, this can be challenging if the circuit is made of non-Clifford gates. To circumvent this issue, protocols like Randomized Compiling (RC) have been proposed \cite{Wallman2016-dn, ferracin2019} and demonstrated \cite{Ware_2021, hashim2020randomized}. In this protocol, the steps are separated between `easy' and `hard cycles' depending on their error rate. The hard cycles usually correspond to the cycles with multi-qubit gates. The twirling gate is inverted through each hard cycle instead of only at the final step. This approach avoids tracking all the twirling gates and the complex inversion gate at the end of the circuit, making it scalable to any number of qubits. However, it requires that any Pauli can be inverted through the entangling gates by another Pauli. It is the case for all the two-qubit gates which are locally-equivalent to Cliffords, like the CNOT and the CZ, but not for entangling gates like the fSim \cite{foxen2020}.


\emph{Pauli Twirling and Cycle Benchmarking}--- Pauli Twirling effectively changes the Pauli-Transfer Matrix (PTM) \cite{Chow_2012} of the error $\Lambda$. It can be shown that the error matrix $\Lambda_{P}$ under Pauli twirling is simply the diagonal of the full PTM $\Lambda$ (\textit{i.e.} the off-diagonal terms of the error are suppressed). This emphasizes the advantage of RC over simply running a circuit: eliminating the off-diagonal terms of $\Lambda$ makes any coherent interference of error impossible and thus increases the algorithms' predictability. Several protocols are specifically designed to measure these Pauli errors \cite{erhard2019characterizing, Harper_2020, Harper_2021, Flammia_2020}. In this article we use Cycle Benchmarking (CB) \cite{erhard2019characterizing} as our main method for measuring the Pauli errors in our system. For each Pauli, CB measures the Pauli decay $\lambda_P$ that corresponds to the diagonal of the $\Lambda_P$ PTM. It can be used to exhaustively measure all Pauli channels - for small sized system - or to statistically sample from a large set of channels.

Even though the number of non-zero entries of $\Lambda_P$ is reduced compared to that of $\Lambda$, there are still $4^N$ elements, making a complete characterization of all the terms not scalable as the number of qubits increases. However, CB demonstrates that sampling from these coefficients can give a good estimate of the behavior under randomized compiling. Introducing $\bar{\lambda}$ as their mean, the Pauli decays are also bounded by (see supplements of \cite{erhard2019characterizing}):

\begin{equation}
    2 \bar{\lambda}-1 \leq \lambda_P \leq 1 , \quad \forall P \in \mathcal{P},
    \label{eq:bound}
\end{equation}

In this article, we argue that the noise under RC with Pauli Twirling can be approximated by a fully depolarizing noise model, where all the Pauli decays except the identity are equal, within a controlled approximation.


\emph{Experimental investigation of RC on random circuits}---  To demonstrate the properties of Pauli twirling under RC, we sample uniformly random 2-qubit circuits (random in SU(4)) and compute the expectation values $E_m$ of all the possible Pauli strings composed of $Z$ and $I$. By compiling the rotations of the eigenbasis of an observable $O$ into the last cycle of the circuit and by removing the identity part of this observable (\textit{i.e.} requiring that $\Tr(O)=0$), we can always map the measurement of $O$ to such a Pauli string. In Figure~\ref{fig:v_shape} we plot the distribution of the mean errors on the expectation value measured $E_m$ compared to its ideal value $E$ as we increase the number of randomizations, thus coming closer to an ideal twirl. We have measured the expectation values of 75 different unitaries for each number of randomizations. To make the comparison fair, we have varied the number of randomly compiled circuits, keeping the total number of shots to 5000. The detailed experimental protocol is described in the supplements. First, the spreading of the errors $|E_m-E|$ is reduced as can be seen on the histograms in Figure~\ref{fig:v_shape}.a. as one would expect from reducing the possibility of error accumulation. Second, in Figure~\ref{fig:v_shape}.b., the mean error without RC is flat as a function of the ideal value, whereas as we increase the number of randomizations, a linear dependency on the ideal value appears, corresponding to a fully depolarizing noise model.

\emph{Estimators for the average depolarization}--- Under a fully depolarizing noise model, the measured expectation value can be written as $E_m = \lambda E$. The scaling factor $\lambda$ can also be understood as the length of the generalized Bloch vector. An estimator of $\lambda$ can be constructed measuring the purity of the state at the end of a circuit. This, however, requires full tomography of the final state. Another estimator is the mean of the Pauli decays $\lambda_P$ measured using CB. As it is a scalable protocol, it is possible to use it to estimate $\lambda$. In Figure~\ref{fig:v_shape} b. we have plotted in dashed lines the average depolarization estimated from the mean of the Pauli decays measured with CB of the CZ, $\bar{\lambda}_{CB} = 0.980$ which shows good agreement. We can extract more information from the CB data: the distance between the two noise matrices $\Lambda_d$ (depolarizing error) and $\Lambda_P$ (twirled errors or Pauli errors) is exactly the standard deviation of the measured Pauli decays under CB. This provides very valuable \textit{a priori} information on how well error mitigation protocols will work, based on the assumption that the error is almost a depolarization model.

\emph{Depolarizing errors can be mitigated by a simple purification}--- A fully depolarizing noise model is simple to mitigate. Knowing $\lambda$, the ideal expectation value can be recovered by rescaling all expectations values by this same coefficient $\lambda$. This technique has been used in Nuclear Magnetic Resonance experiments and more recently in \cite{vovrosh2021efficient}. In Figure~\ref{fig:v_shape} c. we have used a full tomography of the 2-qubit state to extract the length of the generalized Bloch vector and then purified the expectations values measured with this information. In this case, the purification is given by: 
\begin{equation}
    \tilde{E_p} = \lambda E_p \quad \text{with} \quad \frac{1}{\lambda^2} = \frac{1}{2^N-1} \sum_{P \in \mathcal{P}^{\otimes n}\setminus \{I^{\otimes n}\}} E_P^2.
    \label{eq:rescaling}
\end{equation}

\begin{figure}
    \centering
    \includegraphics{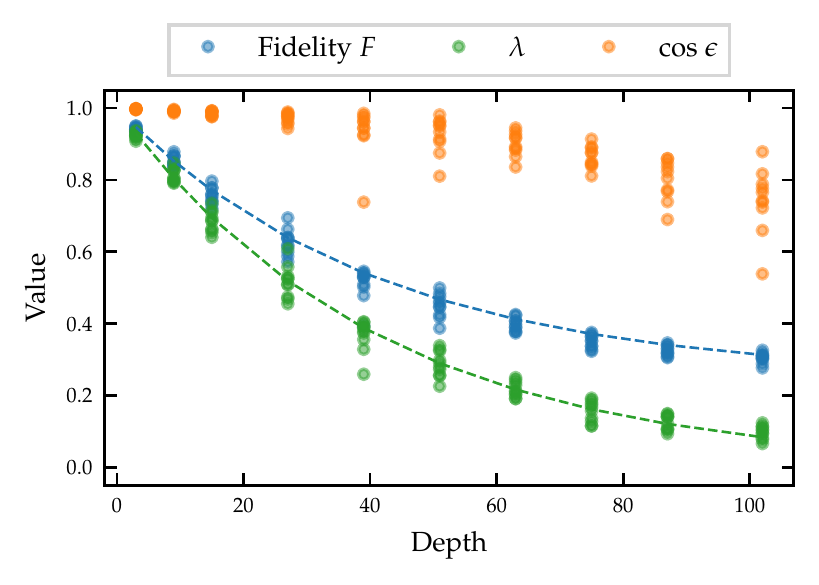}
    \caption{Fidelity $F$ of random states versus the number of CZ gates presented in blue points, together with the independent estimation from CB, using equation (\ref{eq:fidelity}) as the dashed blue line. When these data were taken, the CB of this CZ was $\bar{\lambda}_{CB} = 0.976$. The length of the generalized Bloch vector $\lambda$, is shown in green, together with its CB estimation in dashed green. The remaining error separating the two contributions corresponds to an angle error as shown by equation (\ref{eq:angle_length}).
    }
    \label{fig:fidelity_vs_depth}
\end{figure}

\begin{figure}
    \begin{center}
    \includegraphics{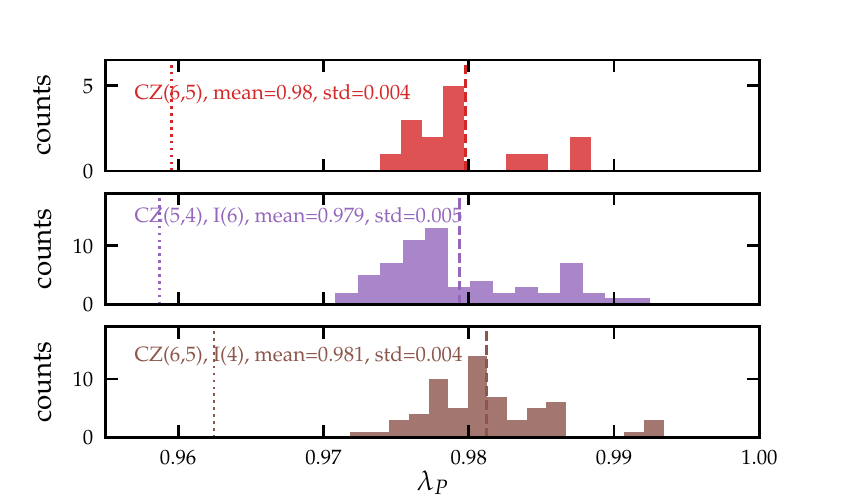}
    \caption{Histograms of the Pauli decays $\lambda_P$ for the different cycles. The first histogram shows the 15 different $\lambda_P$ for the CZ gate of Figure \ref{fig:v_shape}. The dashed line indicates the mean and the dotted line the lower bound of equation (\ref{eq:bound}). The 2 other histograms show for comparison the corresponding parameters when characterizing the different 3-qubits CZ hard cycles, including a remaining idling qubit. The spread of the 63 corresponding $\lambda_P$ is shown for the 2 cycles used in the following QITE experiments.
    }
    \label{fig:lambdaP_2}
    \end{center}
\end{figure}

\emph{Increasing the depth}--- In Figure~\ref{fig:fidelity_vs_depth}, we have increased the depth of two-qubit circuits. For two qubits there is only one hard cycle, so the depth is the number of times this hard cycle is applied. The sampling is done by picking random unitaries from $SU(4)$ and using a KAK decomposition. For a depth-6 circuit for example, we have randomly chosen two unitaries, decomposed them into our native gate set and then measured the concatenated circuits. We use this method to preserve the uniform sampling for all the different depths. The fidelity of a measured state $\rho$ to a pure state $\sigma$ can be expressed in terms of their Pauli expectation values:

\begin{align}
    F(\rho,\sigma) & = \tr(\rho \sigma) \\
      & = \frac{1}{2^N} \Big(1 +\sum_{P \neq I^{\otimes N}} \rho_P \sigma_P \Big) \\
      & = \frac{1}{2^N}  + \left(1- \frac{1}{2^N} \right)\lambda \cos \epsilon
    \label{eq:angle_length}
\end{align}

where $\lambda$ is the length of the Bloch vector given by Equation (\ref{eq:rescaling}) and $\epsilon$ is the angle between the two generalized Bloch vectors. Equation (\ref{eq:angle_length}) indicates that two mechanisms can decrease the fidelity: a reduction of the Bloch vector length $\lambda$ and a misalignment of the vectors axes giving a contribution $\cos \epsilon$. Purification or re-scaling is intended to correct the first type of error, but leaves untouched the second kind of error, or angle error. Assuming that the first kind of error is the limiting one, the fidelity can be approximated as
\begin{align}
    F(\rho,\sigma) \simeq \frac{1}{2^N}  + \left(1- \frac{1}{2^N} \right)\lambda.
    \label{eq:angle_length_approx}
\end{align}
For a circuit with the same hard cycle repeated $n_{\text{cycles}}$ time and with an average Pauli decays $\bar{\lambda}$ we can simply write that:
\begin{equation}
    F(\rho,\sigma,n_{\text{cycles}}) \simeq \frac{1}{2^N}  + \left(1- \frac{1}{2^N} \right)\bar{\lambda}^{n_{\text{cycles}}}.
\label{eq:fidelity}
\end{equation}

The measured fidelities obtained from full tomography for 10 different depths and 10 random circuits at each depth, randomly compiled 20 times are shown in Figure~\ref{fig:fidelity_vs_depth}. The estimation of equation (\ref{eq:fidelity}) using $\bar{\lambda}$ from CB in dashed blue shows a very good agreement indicating that the length errors are indeed dominating. We also extract the length $\lambda$ for each random circuit, and compare it to $\bar{\lambda}^{n_{\text{cycles}}}$. We also extract the residual error  using the following equation:
\begin{equation}
    \cos{\epsilon} = \frac{1}{2^N-1} \sum_{P \neq I^{\otimes N}} \frac{\rho_P}{\lambda} \sigma_P.
\end{equation}
This part of the error cannot be corrected with a simple purification. We note that the degradation of the fidelity due to the angle error is much slower than the part due to reduction of the Bloch vector length and that with purification, circuits with larger depth can be explored. This result emphasises that randomized compilation makes circuit performance much more predictable, and that CB is a good tool for predicting circuit performance under RC. 

\emph{Increasing the number of qubits}--- When increasing the number of qubits, several hard cycles need to be considered. In our case, for a linear topology of 3 qubits, we need to consider the 3-qubit Pauli decays obtained by CB for the entangling gates between each qubit. Under randomized compiling the effective rescaling factor $\lambda$ will be given by the product of the $\lambda_i$ obtained for each hard cycle: $\lambda_{\text{eff}} = \lambda_1^{n_1}\lambda_2^{n_2}$ with $n_i$ the number of occurrence of the hard cycle. We note that universal circuits can be constructed using few different hard cycles and single qubit gates, reducing the number of hard cycles it is necessary to characterize. In Figure~\ref{fig:lambdaP_2}, the spread of the Pauli decays $\lambda_P$ are shown for the CZ gate used in Figures~\ref{fig:v_shape} and \ref{fig:fidelity_vs_depth}, and for the CZ cycles of the 3-qubits. We also emphasize that the number of Pauli decays needed to benchmark isolated two-qubit gates (15 for CZ(6, 5)) is much less than the number required to benchmark larger cycles containing idling spectator qubits (64 for CZ(5, 4) in parallel with I(6)). We emphasize that exhaustive sampling of the Pauli decays is not needed as the mean and standard deviation can be estimated efficiently by randomly sampling the Pauli decays \cite{erhard2019characterizing}. We notice that the bound from equation (\ref{eq:bound}) is indeed valid for these data.

\section{Application to the QITE algorithm} \label{sec:qite}

Imaginary time evolution is a classical iterative algorithm to find the ground state of an Hamiltonian. The key ingredient of this algorithm is that the imaginary time propagator $U(\beta) = \exp(-\beta \mathcal{H})$ --- which is non-unitary --- will converge to the ground state for large imaginary time, given that the initial state overlaps with the ground state \cite{adhikari2000numerical,auer2001fourth}. In \cite{motta2020determining}, the authors describe how to use a quantum computer to perform the imaginary time evolution on NISQ hardware without ancilla qubits. The main idea is to normalize the evolution operator at every time step to make it a unitary evolution that can then be decomposed into gates. This can be done efficiently by simply solving a linear system, which is an easy task for classical computers, and thus QITE is free of the complex optimizations that arise in the VQE scheme \cite{lavrijsen2020classical}. The price to pay is that this algorithm is not a fixed depth circuit, but rather will increase the number of gates for every iteration. Recent experimental and theoretical works try to minimize this issue by aggressively reducing the number of steps needed to reach the ground state \cite{yeter2019practical} or by compressing all the steps into a shorter circuit \cite{Lin2020-zd, Gomes2020-gz}. 

\emph{TFIM model}--- In this work, we concentrate on the Transverse Field Ising Model (TFIM). This is a very well known model which has been investigated several times with the QITE and other algorithms \cite{motta2020determining, Sun2020-jw,cervera2018exact,aydeniz2021scattering}, thus proving a proper benchmark for our error mitigation scheme. The TFIM Hamiltonian for a chain of $N$ qubits is:
\begin{equation}
   \mathcal{H} = J\sum_{\langle ij \rangle} X_{i} X_{j} + h\sum_{i} Z_{i}
   \label{eq:ising}
\end{equation}
where $J$ is the interaction exchange between the nearest neighbors, $h$ is the transverse field applied to the chain and $\langle ij \rangle$ indicates that the sum is over nearest neighbors. The state and the evolution operator at a given step can be written on the Pauli Basis:
\begin{equation}
   \rho = \sum_{P \in \mathcal{P}^{\otimes N}} \rho_P P \quad \text{and} \quad U = \exp\left(-i\sum_{P \in \mathcal{P}^{\otimes N}} a_P P\right),
\end{equation}
where $\rho_P$ are the expectation values of the Paulis of the state $\rho$ and $a_P$ are the generators of the unitary $U$. We call support of the state the Paulis that have non-zero expectation values. The TFIM Hamiltonian presents several symmetries that allow to reduce the problem from two perspectives: the construction of the unitary and the support of the ground state (described in Supplements). In Motta \textit{et al} \cite{motta2020determining}, a domain size $D$ is introduced, which can be smaller than the full domain considered by the Hamiltonian, and the Trotterization happens over the different small domains $D$. This will be mandatory for bigger systems, but for the small systems considered here we will use $D$ of the same size as the number of sites of the Hamiltonian.

\emph{Unitary and circuit construction}--- In this section we discuss how we have synthesized each imaginary time step. The generators $a_P$ are calculated via linear regression as discussed in \cite{motta2020determining}. The number of Pauli expectations values to measure as well as the number of generators depend on the size $D$ of the domain considered in the QITE experiment. As our current processor has a limited number of qubits, we have chosen to consider a domain size equal to the total size of the system. This choice allows us to compress the circuit at every step and avoid stacking the gates. It enforces, however, a measurement of the state on its full support and makes the synthesis harder as it uses multi-qubit gates with more than 2 qubits. Using a state-of-the-art circuit synthesis algorithm, QSEARCH \cite{mark2020}, we were able to run all of the 3-qubit TFIM QITE steps with circuits containing less than 12 CZs (see Supplements). We note that enforcing the symmetries on the unitary allowed us to drastically reduce the number of entangling gates. For larger qubit numbers the current approach will have to be improved, but we also expect that in the future, the synthesizers will continue to become more efficient. We also expect that the synthesis could be further tailored for the QITE algorithm. This type of synthesis has recently allowed to find fixed depth ansätze for some iterative algorithms and specific hamiltonians \cite{bassman2021}.

\begin{figure}
        \begin{center}
          \includegraphics{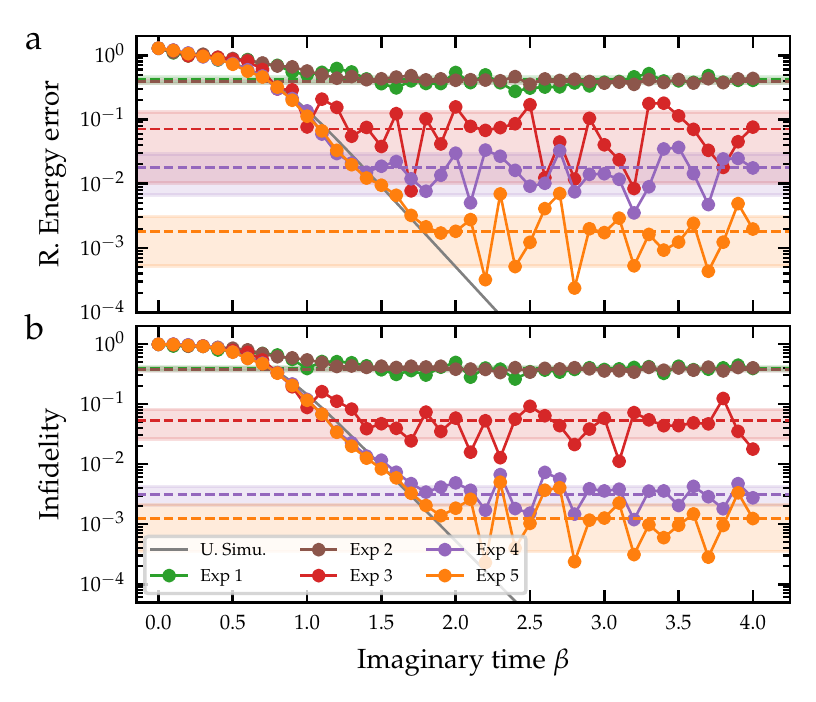}
            \caption{QITE trajectories of relative energy error and infidelity to the ground state for the 3-qubit Ising model, with $J=h=1$. The number of shots and of randomized compilations per point is varied. The mean and standard deviation of the 10 last points are shown by a dashed line and a colored zone. The different colors correspond to:
            U. Simu.: Unitary Simulation;
            Exp 1: No purification, no RC, 20000 shots;
            Exp 2: No Purification, 20 RC, 1000 shots;
            Exp 3: Purification, no RC, 20000 shots;
            Exp 4: Purification, 20 RC, 1000 shots;
            Exp 5: Purification, 20 RC, 1000 shots, and McWeeny purification.
            }
        \label{fig:qite_trajectory}
        \end{center}
    \end{figure}

\emph{Results}--- In order to showcase the error mitigation developed in the first section of this letter, we have run the QITE algorithm on the 3-qubit TFIM for several sets of parameters. In Figure \ref{fig:qite_trajectory}, we plot the QITE trajectory for the parameters $J=h=1$ and calculate both the relative energy error and the infidelity of the measured ground state. For each QITE trajectory we varied the number of RC but keeping the total number of shots constant to show the effect of the different methods independently (Exp 1, 2 and 3). This shows that our method mitigates the jumps previously seen with this algorithm. We indicate on the plots the average of the last points as well as the standard deviations of the measurement after the ground state is reached. This is done in order to capture the stability of the algorithm when the ground state is found. On this plot we see that without randomized compiling - or error tailoring - nor error mitigation, the results are far away from the ground state. Using a purification technique improves the result significantly as it diminishes the impact of incoherent errors introduced by noise tailoring via RC. We note here that iterative algorithms like QITE are particularly sensitive to errors on the expectation value as these values are necessary to determine the next circuit. We then have used both randomized compiling and the purification technique described in the previous section (Exp 4). As we can see the combination of both the noise tailoring and the error mitigation greatly improves the results. We further use the McWeeny purification \cite{mcweeny1960some}, also used for example in \cite{google2020hartree}, corresponding to an iterative procedure projecting the measured state to the closest state of purity one (Exp 5). As we increase the number of randomization and the number of shots per randomization, we are able to push the precision to 0.2\% for both the energy and the ground state infidelity. It is also possible to compute the first excited state, as described in the supplements.
    

\begin{figure}
        \begin{center}
        \includegraphics{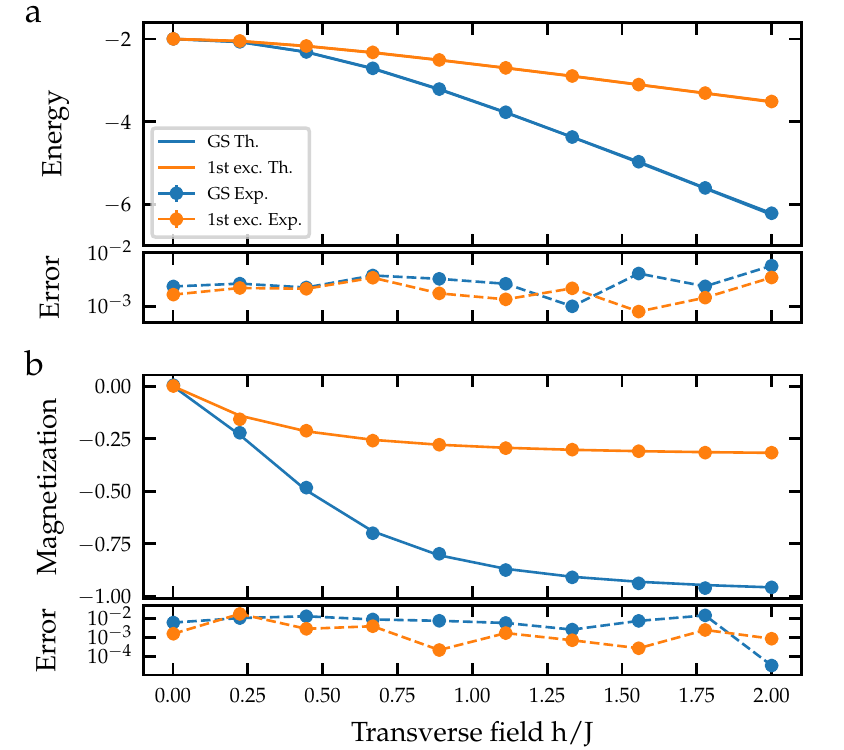}
        \caption{Phase diagram for the 3-qubit Ising model varying the $h$ parameter. a. Energy in units of $E/J$, for the ground state and first excited state. The experiments use 10 randomized circuits and 1024 shots for each value of $h$. The reported energy corresponds to the mean of the 5 last points of the QITE evolution. Error bars corresponding to the standard deviation on these points are smaller than the markers. The error shown for each point corresponds to the relative error. b. measured magnetization of the same data, with the corresponding absolute error. 
        }
        \label{fig:phase_diagram}
        \end{center}
    \end{figure}

    \emph{Phase diagram}--- We then proceed to measure the phase diagram of the TFIM on 3 qubits as the external magnetic field $h$ is swept. In Figure \ref{fig:phase_diagram} we plot the energies and the local magnetization as a function of the transverse field $h$ for both the ground state and the first excited state. For both these quantity, we have used the average over the last states as depicted in Figure \ref{fig:qite_trajectory}. We consistently get errors below 1\% for both the energy and magnetization. 
    
\section{Conclusions} \label{sec:conclusions}
In this letter, we have demonstrated that tailoring the errors with Randomized Compiling simplifies the noise process present on our quantum processor to a Pauli model. We also have shown that this error process is approximately a fully depolarizing noise. We have quantified the distance of our noise process from a fully depolarizing noise channel using the standard deviation of the Pauli decays obtained through Cycle Benchmarking. With this measure, we are able to predict whether simple purification techniques that compensate for this full depolarization can give an advantage. Since the spread of these Pauli decays is bounded, we believe that as the number of qubits increases, there will always be a regime where the noise under Pauli twirling can be approximated by a depolarizating model. In this work, we have concentrated on randomized compiling with Pauli twirling. The spread of Pauli decays depends on the choice of twirling group. The Pauli twirling gives the maximum spread of Pauli decays with a different value for each, but is straightforward to implement. The Clifford group would lead to no spread at all but is impracticable for more than 2 qubits. The Diehedral twirl \cite{carigan2015} can be used with our native CZ gate and will reduce the spread as Pauli decays as the ones in $X$ and $Y$ will share the same values. This would reduce the distance to the depolarization channel and therefore potentially enhance the performance of purification techniques. We also foresee that using smart synthesizers that can exclude subspaces with large Pauli decays would help improve the performance of purification techniques. 

We have demonstrated how an iterative algorithm --- Quantum Imaginary Time Evolution --- can benefit from the application of both a noise tailoring technique like randomized compiling and an error mitigation technique like purification. The application of RC and purification results in an improvement over each technique used separately. The approximation made can be tested by measuring the spread of the Pauli decays using CB. In this article, we have concentrated on a simple purification scheme. However, it can be combined with sophisticated techniques, such as error extrapolation methods \cite{kandala2019error} or symmetric post-selection \cite{mcardle2019, bonet2018} to even further enhance the accuracy of the results.

\section*{Acknowledgements} \label{sec:acknowledgements}

This work was supported by the Quantum Testbed Program of the Advanced Scientific Computing Research for Basic Energy Sciences program, Office of Science of the U.S. Department of Energy under Contract No. DE-AC02-05CH11231 and through the Office of Advanced Scientific Computing Research Accelerated Research for Quantum Computing Programs.

\section*{Competing interests}
I.H, J.J.W., and J.E. have a financial interest in Quantum Benchmark Inc. and the use of True-Q software. The other authors declare no competing interests.

During the writing of this manuscript, we became aware of comparable methods used in \cite{urbanek2021mitigating}.

\bibliographystyle{apsrev4-1}
\bibliography{references}

\clearpage
\section*{Supplements} \label{sec:appendix}

\subsection{Noise description}

\emph{Generic Noise}--- A good way to understand how twirling works is considering the so-called Pauli-Transfer Matrix (PTM) representation of quantum operators. The PTM is simply a matrix representation, in the Pauli basis, of the linear transformation on the density matrix. Hence, a PTM is a square matrix of size $4^N \times 4^N$, where $N$ is the number of qubits. When considering a noisy implementation  $\tilde{A}$ of an ideal operation $A$, we can define the noisy part as $\Lambda = \tilde{A}A^{-1}$. For an ideal implementation, $\tilde{A}=A$ and  $\Lambda$ is the identity. In general, this error matrix can have terms on almost every position. Methods like process tomography \cite{Chuang_1997} and Gate Set Tomography \cite{Blume_Kohout_2017} allow one to reconstruct this matrix using experimental measurements. The off-diagonal terms are specifically harmful for algorithms and prediction of performance as most of them will lead to accumulation of coherent errors. It can be visualized in the single-qubit case with a simple over-rotation: if the error is small for a single gate, repeating the same gate several times makes the error grow quadratically with the circuit depth. Additionally, the error is dependent on the circuit layout, making it hard to predict how well a given circuit will perform. This is one explanation for the gap between actual circuit performance results and what RB would predict \cite{proctor2020measuring}.

\subsection{Variance of the expectation values errors}

\emph{Measuring an expectation value}---
Most algorithms use quantum hardware to measure some specific expectation values that are relevant for the calculus of some energy or to determine the wave-function. In this report we will concentrate on the expectation value of Pauli operators. Following \cite{erhard2019characterizing}, we define our estimator as following. For a given $N$-qubits Pauli operator $Q$, let $\mathcal{B}_Q$ be the rotation that maps the computational basis to an eigenbasis of $Q$ (\textit{eg} a $R_y(\pi/2)$ for the single-qubit operator $X$). The measurement protocol gives an outcome $z$ after the rotation in the eigenbasis with $\mathcal{B}^\dagger_Q$. The expectation value of $Q$ can then be expressed as:
\begin{equation}
    \Tr\left[Q\rho\right] = \sum_{z \in \mathbb{Z}^N_2} \Tr\left[\mathcal{B}_Q(\ketbra{z}{z})Q\right] \Pr(z \vert Q)
\end{equation}
which amounts to a weighted sum over the population measurement.
Due to the discrete nature of the measurement (each single-shot yield a bit string), one need to repeat several times the same circuit (or sequence) and average in order to have an estimate of the expectations value.

\emph{Theoretical estimation of the variance of the errors}---
For a given sequence, the variance evolves as:
\begin{equation}
    \text{Var} \left[E\right] = \frac{\left( 1-E \right)\left( 1+E \right)}{N_{shot}}
\end{equation}
where $N_{shot}$ is the number of repetitions of the measurement and $E$ is the expectation value. This variance is maximal for $E=0$ and minimal for $\vert E \vert=1$. Usually, we experimentally go to large enough $N_{shot}$ to eliminate this source of noise.

\begin{figure}
    \includegraphics[width=14cm]{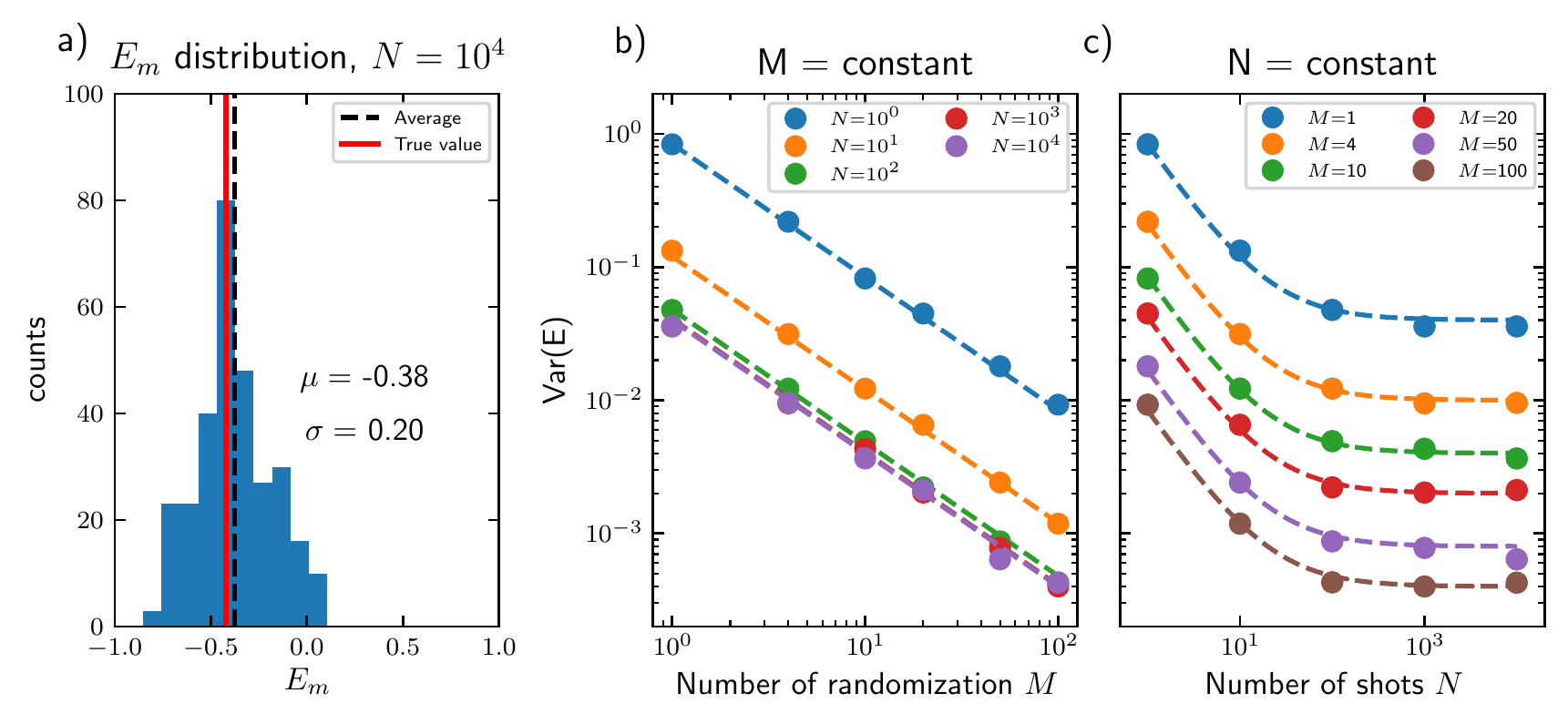}
    \caption{Study of the distribution of the measured expectation value for Randomly compiled circuits of the same unitary. a) Dispersion of the single circuit estimator $E_m$ in the limit of $N \gg 1$ so the single-shot noise is negligible. From this distribution, we extract the variance $\sigma$ of the results from random implementation to another one. b) presents the evolution of the variance of the Randomized compiled estimator $E_{RC}$ as the number of randomization increases. For each curve, the number of shot per random circuit is $N$. c) presents the same variance as a function of the number of single-shots $N$ per randomization. For large number of shot, the variance plateau to a value given by $\sigma^2/M$ as expected from \eqref{eq:variance}. For both figure b) and c), the dashed lines are models following the equation \eqref{eq:variance} and the extracted value from the distribution of a).
    }
    \label{fig:shotnoise}
\end{figure}

Even though the Variance is reduced by increasing the total number of shots, it is not clear that the measured expectation value is the right one. Actually, due to noisy gates, this estimator is biased. This bias depends strongly on the given implementation of the circuit. One way to construct a less biased estimator is to randomly sample from a set of circuits that realises the same unitary \textit{i.e.} the total operation is identical but the circuits differ. Randomized Compiling \cite{Wallman2016-dn} - or Pauli Twirling - gives a framework to do this. The implementation of the same operation only differ by the insertion of Pauli-gate. The estimator for a given expectation value is then given by the average over all the implementations:
\begin{equation}
    E_{RC} = \frac{1}{M} \sum_{m=1}^{M} E_m    
\end{equation}
where each $E_m$ correspond to a different circuit. The advantage of using this approach is that the bias of the $E_{RC}$ estimator is much more predictable, allowing to compensate for it. Under this assumption, the variance of the $E_{RC}$ estimator is given by \cite{Hincks2018-zv}:
\begin{equation}
    \text{Var} \left[E_{RC}\right] = \frac{1}{M}\left(\frac{\left( 1-E_{RC} \right)\left(1+E_{RC} \right)}{N}+ \frac{N-1}{N} \sigma^2 \right) 
    \label{eq:variance}
\end{equation}
As pointed out in \cite{Hincks2018-zv}, the main information from this equation is that the total variance goes to zero as the number of circuits increase, but asymptotes to the finite value $\sigma^2/M$ if the number of randomization is fixed and we increase the number of single-shot $N$. We have probed this with an experimental dataset in Figure \ref{fig:shotnoise}. When choosing between more random circuits or more single-shot, the best choice is to use more random circuits, with the best situation being one shot per random circuits ($N=1$). However, from an experimental perspective, it is not time efficient to run one single-shot per random circuits as the upload time to the control hardware is time consuming whereas repeating the same random sequence is fast. From this experiment, we can see that using up to a few hundred of single-shots per random circuits gives a good compromise. We can also see that getting more than $10^3$ single shots doesn't improve the variance of the result for a fixed number of randomizations. This is of practical use when designing the RC sequence for a given circuit.

\onecolumngrid\
\twocolumngrid

\subsection{Detailed experimental protocol for Figure~\ref{fig:v_shape}}

In practical settings, one is interested in measuring expectation values. To study the effect of Randomized compiling when measuring expectation values we have run a series of tomography performed as following:
\begin{enumerate}
    \item Draw a random unitary from a uniform sampling of SU(4).
    \item Decompose the unitary into a circuit using the KAK decomposition (3 CZs). Repeat until reaching the wanted depth (6 CZs for Figure~\ref{fig:v_shape}).
    \item Create 9 circuits by appending the measurement rotation at the end of the circuit.
    \item For each circuit generate M Randomly compiled circuits.
    \item Run each of the $9 \times M$ circuits $N$ time to gather statistics.
    \item Calculate the expectation value for each Pauli for each circuit (15 per unitary).
\end{enumerate}

\subsection{Unitary and circuit construction}
    We discuss here how we have synthesized each imaginary time step step. The generators $a_P$ are calculated via linear regression as discussed in \cite{motta2020determining}. When performing this linear regression, the result is very dependent on the measurement noise. To mitigate this effect, we used a Ridge regression that constrained the overall norm of the generators. The number of Pauli expectations values to measure as well as the number of generators depend on the size $D$ of the domain considered in the QITE experiment. As our current processor has a limited number of qubits, we have chosen to consider only domain sizes equal to the total size of the system. This choice allows us to compress the circuit at every step and avoid stacking the gates. It enforces, however, a measurement of the state on its full support and makes the synthesis harder as it uses multi-qubit gates with more than 2 qubits. Using a state-of-the-art circuit synthesis algorithm, QSEARCH \cite{mark2020}, we were able to run all of the 3-qubit TFIM QITE steps with circuits containing less than 12 CZs. We note that enforcing the symmetries on the unitary allowed us to drastically reduce the number of entangling gates. We report in table \ref{tab:number_cnots} the number of entangling gates for different synthesizers. QSEARCH allows to set the numerical accuracy of the synthesis of the unitary, and it is possible to improve the computational time by reducing the accuracy while still not limiting the ground state fidelity measured on the hardware. We note that for larger qubit numbers the current approach will have to be improved, but we also expect that in the future, the synthesizers will continue to become more efficient. We also expect that the synthesis could be further tailored for the QITE algorithm. We note that this type of synthesis has recently allowed to find fixed depth ansätze for some iterative algorithms and specific hamiltonians \cite{bassman2021}.

    \begin{table}
    \begin{center}
        \begin{tabular}{ |c|c|c|c| } 
         \hline
         N & QISKIT & QFAST & QSEARCH \\
         \hline
         2 & 3 & 3 & 3 \\ 
         \hline
         3 & 30-35 & 10-12 & 7-12 \\ 
         \hline
         4 & 160-200 & 70-80 & 30-50 \\ 
         \hline
        \end{tabular}
        \caption{Comparison of the maximum number of entangling gates obtained with different synthesizer. QSEARCH \cite{mark2020} and QFAST \cite{younis2020qfast} and the generic tool Isometry on Qiskit \cite{Qiskit} for a QITE simulation of a TFIM of $N$ sites with the parameters $J=h=1$.}
        \label{tab:number_cnots}
    \end{center}
    \end{table}
    

\subsection{Symmetries of the Hamiltonians}

We describe here how several symmetries of the TFIM Hamiltonian have been used to reduce the number measurements needed, and of generators. This can apply to many Hamiltonians.

    \emph{$\mathbb{Z}_2$ symmetry}--- Let's first consider the so-called $\mathbb{Z}_2$ symmetry: The Hamiltonian from equation (\ref{eq:ising}) commutes with the operator $Z^{\otimes n}$. This symmetry divides the full Hilbert space into two eigenspaces with, for any state $\ket{S}$, $Z^{\otimes n} \ket{S} = + \ket{S}$ or $Z^{\otimes n} \ket{S} = - \ket{S}$. All the eigenstates of $\mathcal{H}$ have to be in either of these subspaces. This property will both restrain the support of the possible ground states and force the QITE evolution operators to preserve the parity with respect to this symmetry. To reduce the support of the ground state, we recall that two Pauli operators can either commute or anti-commute. For the Pauli operators that anticommute with the symmetry $S$, the expectation value of this Pauli on an eigenstate of the Hamiltonian is necessarily zero: $\expval{P}{\text{GS}}= \Tr(\rho P) =  \rho_P = 0$. If we require all the steps to fall within the symmetric subspace, this will also enforce that $U$ commutes with the symmetry $S$. Developing the evolution to the first order, we find that $U$ commutes with $S$ if and only if $ a_P = 0~\forall P \in \mathcal{P} ~\text{such that}~\acomm{P}{S}=0$. This simplifies the synthesis as the number of generators to consider is reduced by a factor two. For the support of the ground state, if we know the sign of the parity $s$, usually found using a classical algorithm, we can further constraint the support by noting that $\expval{SP} = s \expval{P}$. The number of free parameters for the ground state is thus reduced by a factor four compared to the full Hilbert space size.

    \emph{Time reversal symmetry}--- As the TFIM Hamiltonian $\mathcal{H}$ is real, it is invariant by Time Reversal Symmetry. This adds another symmetry to consider. The corresponding symmetry operator is $T = K$, with $K$ the complex conjugation. The unitary evolution at each step has to commute with $K$, which implies that the generator has to anti-commute. This means that the support of the generators is included in the Pauli matrices that anti-commute with $K$: $ a_P = 0~\forall P \in \mathcal{P} ~\text{such that}~\comm{P}{K}=0$. This implies that the set of allowed generators are the Pauli strings with an odd number of $Y$. It also means that the ground state should be an eigenvector of $K$ meaning that the support of the eigenstate has to commute with $K$. In other words, the eigenstate support set is intersected by the Pauli strings that have a even number of $Y$. We note that for this specific symmetry, the set of generators and the set of support for the ground state are disjoint.

\subsection{First excited state and higher energies levels with the QITE algorithm}

The QITE algorithm offers an efficient way to calculate the higher energy states. When the ground state $\ket{\text{GS}}$ is determined without considering the symmetries, one idea is to add to the Hamiltonian $\mathcal{H}$ a term proportional to the ground state in order to make the first excited into a ground state: $\mathcal{H} \rightarrow \mathcal{H} + \alpha \dyad{\text{GS}}{\text{GS}}$ with a coefficient $\alpha$ large enough. It is also possible, for the first excited state, to make use of the symmetry: the ground and first excited states should have opposite parities. Then using the same QITE algorithm, but changing the initial state parity allows to find the 1st excited state. This is what was done for Figure \ref{fig:phase_diagram}.

\end{document}